# Quantum phase transition of the $^{103m}$Rh spin-density wave

Y. Cheng

*Department of Engineering Physics, Tsinghua University, Beijing, 100084, China*

Abstract

We induce a quantum phase transition of the $^{103m}$Rh excitation at the critical density of $10^{12}$ cm$^{-3}$ by bremsstrahlung pumping at 300 K. A massive $^{103m}$Rh spin-density wave carrying a spin current moves on the identical $^{103}$Rh matrix like a quantum fluid. The collapse-and-revival spectral evolution indicates that the collective nuclear excitation undergoes dynamic Bose-Einstein condensation. Applying an external magnetic field, we observe the Onsager-Feynman quantization, which gives further evidence of the superfluidic phase.

## 1. Introduction

$^{103m}$Rh, a long-lived nuclear state at 39.76 keV with a 56 minute half-life, is excited by bremsstrahlung pumping. The emission spectra of $^{103m}$Rh have been investigated in earlier works using time-resolved gamma spectroscopy at 300 K and at 77 K [1-3]. With the low inversion density of the two-level Mössbauer nuclei, the luminosity at 300 K is proportional to the gamma irradiation exposure (regime I), but it increases nonlinearly at higher exposures (regime II) [1, 2]. Further increasing the exposure puts $^{103m}$Rh into regime III, where the luminosity no longer increases but decreases.

The spectral profiles of the K X-rays and the γ rays are slightly broadened in regime I [1]. In regime II, the spectral profiles are significantly broadened, and the inversion density increases nonlinearly [1]. A γ ray triplet profile appears when the splitting energy exceeds the detector resolution (about 400 eV). When the sample is cooled from 300 K to 77 K, the broadened triplet in regime II abruptly contracts, exhibiting a spectral collapse-and-revival [2]. The same spectral contraction and spectral collapse-and-revival also appear in regime III at 300 K. The temperature-dependent critical density suggests that the collective excitation undergoes a phase transition.

We observe anomalous emissions (AE) centering around half the transition energy ($E_M/2$ = 19.9 keV), and will report it elsewhere. The AE from $^{45m}$Sc excited by bremsstrahlung pumping [3,4] and from $^{93m}$Nb excited by neutrons is also observed in our preliminary studies, whereas AE is absent in the 52 % - 48 % mixture of $^{107m}$Ag and $^{109m}$Ag. The three nuclides Rh, Nb, and Sc are all present as single isotopes with 100% natural abundance. These findings suggest that the collective channel opens for the identical nuclei in crystals.

Gamma rays incident at the Bragg angle give rise to a collectively enhanced reflection from crystals where the photo-electric effect is suppressed due to the vanishing E-field at the lattice sites. Electronic scattering is suppressed for the two γ beams propagating in the Borrmann mode [5], whereas in crystals these two γ beams strongly couple with the identical multipolar transition. Cascade decay recoilless channels open at the Bragg angle for the photon pair simultaneously emitted as an entangled biphoton [6, 7].

Recoilless and transparent channels also open with Brillouin scattering [8]. Entering the strong-coupling regime [7], the loss of the energy and the momentum due to the biphoton emission via the Stokes process is immediately recovered by the anti-Stokes process, thereby conserving energy and momentum by exchanging the biphoton with the phonon. The small Stokes and anti-Stokes shifts of the biphoton lead to the observed AE peaking around $E_M/2$; i.e., at 19.9 keV for $^{103m}$Rh, 6.2 keV for $^{45m}$Sc, and 15.4 keV for $^{93m}$Nb. Analogous to the plasma mode screened by electrons, $^{103m}$Rh is screened by the biphoton field at resonant nuclei forming a nuclear spin-density wave (NSDW), also called a nuclear exciton [5]. Via the

Anderson-Higgs mechanism the NSDW conspires with the longitudinal phonon to acquire an eV mass [9].

The hard-core neutral quasiparticle carrying a spin current travels over the $^{103}$Rh matrix like a quantum fluid. Due to the light eV mass, NSDWs exhibit a giant magneton on the order of meV/gauss. The strong end-to-end interaction between NSDWs forms 1D spin chains with a 2D antiferromagnetic ordering due to their bosonic exchange. The anisotropic emission of the NSDW texture depends on the macroscopic sample geometry, revealing that rotational symmetry is spontaneously broken. Beyond a critical density, massive quasiparticles undergo dynamic Bose-Einstein condensation (BEC) [10]. Recently, the Cooper pairing of Luttinger liquid with bosonic atoms has been predicted [11]. In fact, the collective coupling between biphoton and nuclei is abruptly reduced upon entering regime III, indicating that the phase transition is most likely triggered by Copper paring with the 4-photon singlet rather than with a biphoton BEC.

The spectral collapse-and-revival reveals the macroscopic coherent condensate observed from ultracold atoms [12]. Observation of the spectral collapse-and-revival constitutes evidence of the NSDW BEC in regime III. Moreover, Onsager-Feynman quantization [13] is recognized by applying external magnetic fields, where the spin of the entire NSDW flips simultaneously. This off-diagonal long-ranged order [14] is further evidence of the quantum phase transition to a superfluidic state.

## 2. Measurements

The experimental procedure, the sample (25 × 25 × 1 mm$^3$), and the detection system configuration are the same as reported previously [1-3]. The detector is horizontally leveled, as shown in figure 1, and oriented along the north-south direction, roughly parallel to Earth's magnetic field. The data acquisition system consists of a digital multichannel analyzer (MCA, CANBERRA DSA 1000). The energy spectra are acquired using a channel width of 25.7 eV.

We measured two different sample orientations, as shown in figure 1. To provide a sufficient counting rate, three samples were sandwiched with two 1 mm thick tungsten sheets. The tungsten sheets prevented stray photons from adjacent samples from entering the detector. The external magnetic field was applied using a Helmholtz coil with 7.62 ± 0.01 gauss/A. The component of Earth's magnetic field parallel to the field applied by the Helmholtz coil was cancelled at 0.04 A, leaving the 0.3 gauss vertical component of Earth's field unchanged.

Figure 2 illustrates the measured luminosity in three regimes as a function of exposure rate. The three data points that fall on the line (filled circles) are in regime I, while the next three data points (one closed circle and two open circles) are in regime II. The seven data points represented by filled triangles are in regime III, and appear in two groups. The first group (numbered 1, 2, and 3) is obtained by increasing the e-beam energy by 2% with respect to the two neighboring (open circle) data points, which were reported in [2]. The fluctuation of the bremsstrahlung intensity is about 1%. The second group on the right-hand side is acquired by further increasing the e-beam energy by 6%. Data points 1, 2, and 3 correspond to experiments with no filter, with a Cu filter, and with a Ta filter, respectively. When the luminosity due to Rh Kα is $10^3$ counts per second (cps), the density of $^{103m}$Rh at the irradiated spot is $10^{12}$ cm$^{-3}$.

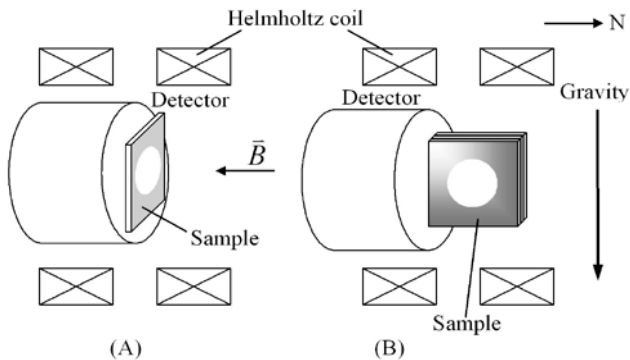

**Fig. 1.** Configurations of the measurements. A Helmholtz coil gives the external magnetic field. The central white spot on the sample stands for the exposure location. Transverse configuration of (A) has short sample axis of 1 mm facing the detector. Longitudinal configuration of (B) has the long sample axis of 25 mm facing the detector. The relative orientation of the set up with respect to the gravity is indicated by the arrow.

## 3. Data analysis

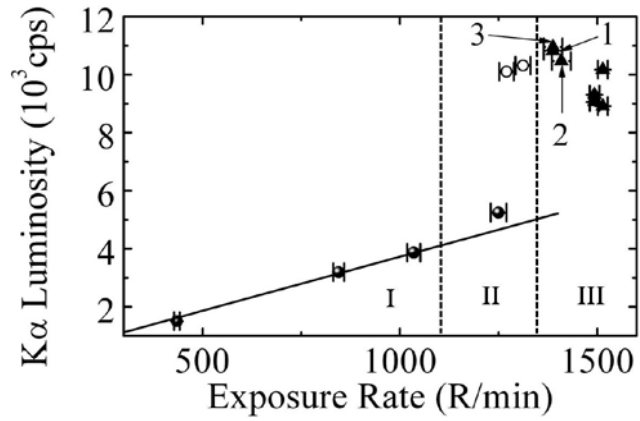

**Fig. 2.** Kα luminosity with configuration (A) at the beginning of measurement in variation with the recorded value of the exposure rate in Roentgen per min (R/min). Three regimes marked with I, II, and III are separated by the vertical dashed lines. Data points with filled circles and open circles are reported in [1] and [2], respectively. Seven data points with filed triangles located in Regimes III show the decreasing luminosity with exposure. Data points indicated by the numbers 1, 2, and 3 are selected for the presentation in this work, without filter, with Cu filter, and with Ta filter, respectively.

Decreasing luminosity in regime III, a crucial property of the system, is linked to several unique behaviors of $^{103m}$Rh NSDWs. First, the spectral deformations, which are defined by subtracting the normal profiles from the measured profiles [1, 2], are no longer stationary but exhibit a collapse-and-revival behavior. The deformation abruptly switches between two typical dynamic patterns, as shown in figure 3. Second, the spectral splittings are not widely spaced, as found in regime II [1, 2], but closely spaced. In particular, the splitting in the spectral shape shown by the blue trace in figure 2 is close to the vacuum splitting observed in regime I [1]. Third, the spectral deformations are less visible when using the filters, and the Ta filter causes a stronger suppression than the Cu filter. Finally, the spectral profiles are static for the second group of data with a further enhanced bremsstrahlung intensity, because the spectral collapse-and-revival immediately occurs after the early phase transition during the irradiation.

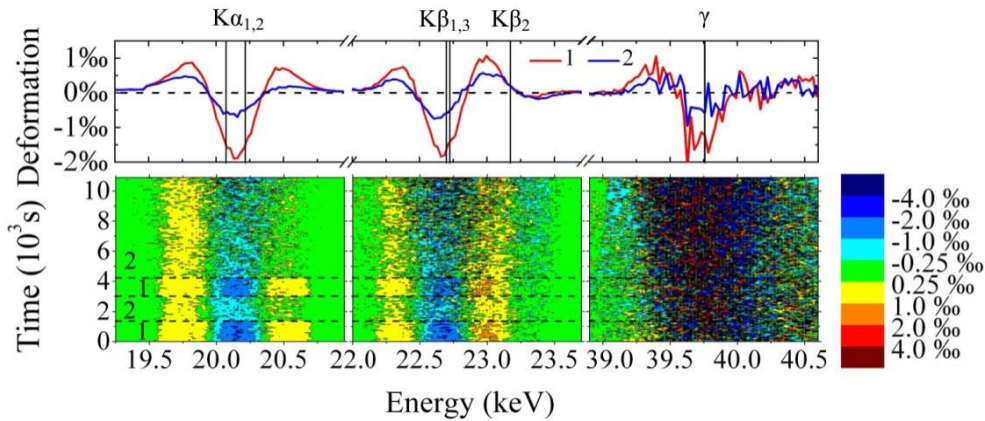

**Fig. 3.** (colour online) The spectral collapse-and-revival in three bands of Kα, Kβ, and γ for the data point 1 without filter, as shown in figure 2. The normalized spectral deformation defined in [1,2] is the measured spectral profile in every minute deviating from the normal profile calibrated for the detector in use. In the lower figure, two different shapes indicated by the number 1 switch back and forth without changing any parameters of detecting system. In the upper figure, line 1 in red color is the deformation of shape 1 and line 2 in blue color is the deformation of shape 2.

Though the irradiation procedure is the same for configurations A and B, as shown in figure 1, the total K counts accumulated in three hours depend on the magnetic field and the macroscopic geometry, as shown in the upper part of figure 4. The gamma emission of the E3 multipolar transition prefers the nuclear polarization along the field. This orienting effect is revealed by a weak applied field of several gauss. This striking field sensitivity reveals the broken rotational symmetry of the NSDW magneton. Furthermore, the anisotropic response to the magnetic field indicates the texture retained by the macroscopic sample geometry. A sawtooth dependence for the photon counts is observed for both configurations, as shown in figure 4 and table 1. Significant dips appear at 0.04 A and 0.5 A, which correspond to horizontal fields of 0.0 and 3.8 gauss, respectively.

The penetration depths of the K lines (about 20 keV) and γ (about 40 keV) are both on the order of 50 μm, meaning that emissions from the sample surface are measured. The intrinsic properties of the $^{103m}$Rh emissions dictated by the atomic structure are constant and isotropic, no mater what configuration is applied. The internal conversion $α_K$, as determined by the K counts per γ count and shown in figure 4, departs the above-mentioned photo-electric attenuation. First, $α_K$ for configuration A is higher than $α_K$ for configuration B, indicating the γ penetration depth is much larger than the 1 mm sample thickness. This penetration depth departs significantly from the photo-electric effect by two orders of magnitude. Second, in figure 4, $α_K$ shows the inverted sawtooth relationship corresponding to the K counts. These two observations indicate that the spectral deformation appearing at the K lines are only a secondary effect induced by the biphoton NSDW.

As shown in figure 4, the orienting effect receives a negligible contribution from the nuclear magneton (on the order of peV/gauss). Taking the NSDW effective mass (~ eV) into account with $\mu = e\hbar/2m^*$, the effective magneton (~ meV/gauss) cannot provide the orienting effect of several gauss at 300 K, unless rotational symmetry is spontaneously broken. The Onsager-Feynman quantization [13] $e\oint v_s dr = eh/m^*$ for the superfluid state is revealed in figure 4 by the dips at ~ 3 gauss with $B = J\, g_l \rho_s \mu_0 eh/2m^*$. The coherent flipping of the NSDW is a manifestation of off-diagonal long-range order. Here, $B$ is the external magnetic induction, $\mu_0$ is the vacuum permeability, $\rho_s$ ~ $10^{12}$ cm$^{-3}$ is the critical density, $v_s$ is the spin superflow velocity, $e$ is the elementary charge, and $h$ is Planck's constant. With the angular momentum $J = 3$ and a proton $g$-factor of $g_l$ ~ 5.6, we estimate an effective mass of $m^*$ ~ eV. We suggest that the nuclear-orienting response is due to the spin Meissner effect [10, 15], and is a consequence of superfluidity [13]. The logarithmic orienting response in figure 4 shows the field-dependence of the magneton.

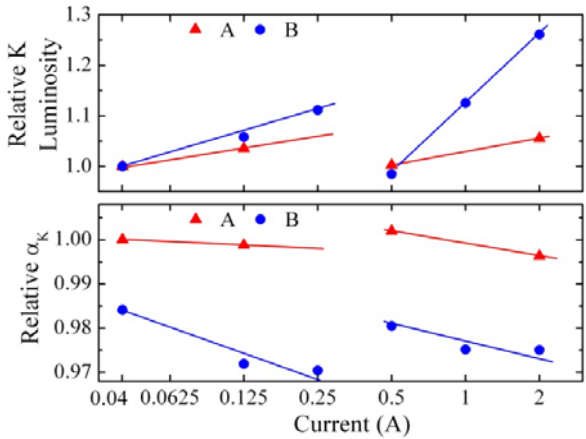

**Fig. 4.** (colour online) The upper figure shows the relative counts of K lines corresponding to the most left data points with configurations A (red triangle) and B (filled blue circle) as a function of the external magnetic field. The lower figure shows the relative conversion ratio $α_K$ between K and γ, where the most left data point of configuration A is set to unity. The correlated variation in photon counts about 6 % with configuration A and about 30% with configuration B naturally manifests itself despites the exposure fluctuation of 1.6%.

**Table 1.** Emissions vs the apply magnetic field. The abbreviations are for K: Rh K lines; γ: γ ray at 39.76 keV.

| Configuration | Emissions | 0.04 A | 0.125 A | 0.25 A | 0.5 A | 1 A | 2 A |
|---|---|---|---|---|---|---|---|
| A | K ($10^4$) | 5524±0.7 | -- | 5717±0.8 | 5538±0.7 | -- | 5830±0.8 |
| A | γ ($10^4$) | 47.97±0.07 | -- | 49.74±0.08 | 48.02±0.07 | -- | 50.84±0.08 |
| B | K ($10^4$) | 219.4±0.1 | 232.1±0.2 | 243.8±0.2 | 216.0±0.2 | 246.9±0.2 | 276.6±0.2 |
| B | γ ($10^4$) | 1.93±0.01 | 2.07±0.02 | 2.17±0.02 | 1.91±0.01 | 2.19±0.02 | 2.46±0.02 |

The author thanks Bing Xia and Xiang Wang for the data preparation, Hong-Jian He and Feng Wu for the fruitful discussions, and Chih-Hao Lee for the $^{93m}$Nb excitation. This work was supported by the NSFC, grant 10675068.